\documentstyle[emulateapj]{article}
\begin{document}

\def\thefootnote{\fnsymbol{footnote}}

\title{The Optical Gravitational Lensing Experiment\\
Monitoring of QSO 2237+0305 \footnotemark}

\author{P.R. Wo\'zniak$^{1}$, C. Alard$^{2}$,}
\author{A. Udalski$^{3}$, M. Szyma\'nski$^{3}$, M. Kubiak$^{3}$,
G. Pietrzy\'nski$^{3}$, and K. \.Zebru\'n$^{3}$.}

\bigskip\bigskip

\affil{$^{1}$Princeton University Observatory, Princeton, NJ 08544--1001, USA}
\affil{e-mail: wozniak@astro.princeton.edu}
\affil{$^{2}$Institute d'Astrophysique de Paris, 98 bis Boulevard Arago,
F-75014 Paris, France}
\affil{e-mail: alard@iap.fr}
\affil{$^{3}$Warsaw University Observatory, Al. Ujazdowskie 4,
00-478 Warszawa, Poland}
\affil{e-mail: udalski,msz,mk,pietrzyn,zebrun@sirius.astrouw.edu.pl}

\bigskip

\begin{abstract}

We present results from 2 years of monitoring of Huchra's lens
(QSO 2237+0305) with the 1.3 m Warsaw telescope on Las Campanas, Chile.
Photometry in the $V$ band was done using a newly developed method for
image subtraction. Reliable subtraction without Fourier division removes
all complexities associated with the presence of a bright lensing galaxy.
With positions of lensed images adopted from HST measurements it is relatively
easy to fit the variable part of the flux in this system, as opposed
to modeling of the underlying galaxy. For the first time we observed smooth
light variation over a period of a few months, which can be naturally
attributed to microlensing. We also describe automated software capable
of real time analysis of the images of QSO 2237+0305. It is expected that
starting from the next observing season in 1999 an alert system will be
implemented for high amplification events (HAE) in this object. Time sampling
and photometric accuracy achieved should be sufficient for early detection
of caustic crossings.

\end{abstract}

\keywords{gravitational lensing: observations -- photometry --
objects: individual: QSO 2237+0305}

\footnotetext{Based on observations obtained with the 1.3 m Warsaw
telescope at the Las Campanas Observatory of the Carnegie
Institution of Washington.}

\section{Introduction}

Optical Gravitational Lensing Experiment (OGLE) is a long term
project focusing on detection and monitoring of microlensing
events. During second phase of the experiment data are collected with
the dedicated 1.3 m Warsaw telescope at the Las Campanas Observatory, Chile
(Udalski, Kubiak and Szyma\'nski 1997). Due to the intrinsically small
probability of the microlensing phenomenon, the main targets are the densest
stellar fields, namely the Galactic Bulge and Magellanic Clouds
(e.g. Paczy\'nski 1996). Some telescope time, however, was also allocated
to observations of other objects in which gravitational lensing is present,
e.g. distant quasars lensed by foreground galaxies. Given good seeing
at the Las Campanas Observatory and the amount of available telescope time,
the OGLE-II survey can provide good time coverage of the variability in
a few selected multiply imaged lenses. Currently two such systems are monitored
by OGLE: QSO 2237+0305 and HE 1104-1805.

Objects of this kind are particularly interesting for cosmology
through determination of time delays. QSO 2237+0305
is a unique system and consists of a high-redshift quasar
at $z=1.695$ quadruply lensed by a relatively nearby galaxy at
$z=0.039$ (Huchra et al. 1985). Because light rays of the four images
pass through the bulge of the foreground galaxy, the optical depth
to lensing on individual stars could be as large as 0.4--0.9 depending
on the image (Webster et al. 1991, Schneider et al. 1988),
making microlensing very likely in Huchra's lens.
Moreover the high degree of symmetry, short distance to the
lensing galaxy, and small image separations produce predicted time delays
of about 1 day or less (Schneider et al. 1988). Thus, intrinsic variability
can be easily distinguished from microlensing effects. Microlensing has already
been discovered in Huchra's lens by Irwin et al. (1989) and confirmed later.
Nevertheless, despite broad interest in this object, so far the best
light curve consists of only 59 measurements in the R band spread over
a period of about 8 years and represents the combined efforts of many observers
(\char31 stensen et al. 1996, Corrigan et al. 1991).

The traditional approach to photometry of this object, i.e. with modeling
of the underlying galaxy light or iterative subtraction of the PSF components
at the positions of the lensed images, requires exceptional seeing and
spatial sampling generally not available with the 1.3 m telescope.
Standard deconvolution algorithms also require superb data and their
uncertainties are not easy to understand. In this paper we present a
qualitatively different approach. Using the optimal image subtraction algorithm
described in detail by Alard and Lupton (1998), and generalized by
Alard (1999) for the case of spatially variable kernel, it is now possible
to match PSFs of two images without noise amplifying division in Fourier space.
Photometry on difference images is free of many complications associated
with other methods.

In Section~2 we describe observations. Section~3 contains a description
of the method with the details of this particular adaptation
and photometry on subtracted images. We summarize the results in Section~4
and future plans in Section~5.

\section{Observations}

All observations presented in this paper were made with the 1.3 m Warsaw
telescope at the Las Campanas Observatory, Chile, which is operated by
the Carnegie Institution of Washington. The ``first generation'' camera
uses a SITe $2048 \times 2048$ CCD detector with $24 \mu m$ pixels resulting
in 0.417 arcsec/pixel scale. Images of QSO 2237+0305 are taken in normal mode
(still frame) at ``medium'' readout speed with the gain 7.1 $e^-$/ADU
and readout noise of 6.3 $e^-$. For the details of the instrumentation setup
we refer to Udalski, Kubiak and Szyma\'nski (1997).

To assure uniformity of the data sets, all observing sequences within
the OGLE project are programmed and can be executed in batch mode. Good
seeing is essential for ground based measurements of Huchra's lens,
independent of the photometric method. Therefore, observations are
restricted to nights with seeing better than about 1.4 arcsec, preferably
with low sky background. Median FWHM of the seeing disk was 1.3 arcsec in
the data presented here. The observing season for QSO 2237+0305 at LCO lasts
from late April to mid December. In satisfactory weather conditions
on a given night two 4 minute exposures in the standard $V$ photometric band
are taken, shifted by a few arc seconds with respect to each other.
This can be done typically once or twice per week without a significant
slow down of the primary program. In 1997, we obtained 80 points during
the observing season. In 1998, QSO 2237+0305 was temporarily removed from
the observing program and there are only 19 points during that period.
However, starting from 1999 we expect to get time coverage comparable
to the 1997 observing season.

The OGLE-II data pipeline automatically detects newly collected frames
and performs bias and flatfield corrections (Udalski, Kubiak and Szyma\'nski
1997). At this point initially reduced frames may be passed to photometric
reductions, provided that there is automated software for that purpose.
We describe such software in Section~3 and 5 along with the planned alert
system for high magnification events in QSO 2237+0305.

\section{Photometry}

\subsection{Image subtraction and difference photometry}

Systems such as Huchra's lens pose a remarkable level of complication for
ground based photometry with medium seeing. Because of small image
separations and their proximity to the galaxy nucleus, each pixel near
the center of the blend contains a light contribution from each of 4 lensed
components. Fainter, but complicated, light distribution from the underlying
barred spiral only makes things worse, with the most significant contamination
due to the galaxy core. In the OGLE data the positions of all the lensed
components fit within an area of about 25 pixels.
Clearly, a full fit of positions and intensities is out of the question,
even assuming that we knew the shape of the galaxy light distribution.
Top panels of Fig.~1 illustrate the observational situation.
On the other hand, this extreme local crowding occurs in the field with very
low density of stars, which complicates determination of the PSF, especially
in the presence of spatial gradients. Therefore the basic strategy adopted
here is to fix as many parameters as possible and avoid fitting the galaxy
light altogether by means of image subtraction. We subtract a reference image
from each of the exposures and obtain a series of frames with only the variable
part of the flux. This should completely remove the foreground galaxy
and leave a blend of four variable components with known geometry,
each of them PSF shaped.

Some preliminary processing is required before one can subtract two images
of the same stellar field. Both images must be resampled onto the same
coordinate grid and PSFs need to be matched. The first step is extracting
a $250 \times 500$ arcsec field centered on the object. In this relatively
small field the spatial gradients of the PSF are not too large for
a low order fit and at the same time the field contains enough stars to obtain
a reliable subtraction. Stars are detected at maxima of the cross-correlation
function with the approximate Gaussian model of the PSF. Cross-correlation
image is done by convolving with the lowered Gaussian filter.
For the coordinate transformation we use a first order fit to coordinates
of about 12 stars found in both a given frame and the reference frame.
A simple algorithm for detecting and removing cosmic rays is run before
resampling of the processed image to the coordinate system of the reference
image. We use a bicubic spline interpolator to perform the resampling.
At this stage a resampled frame and the reference frame are fed to the main
program which calculates spatially variable convolution kernel between
both frames. This code utilizes a newly developed method by Alard and Lupton
(1998). Further development of the method for spatially variable kernels
(Alard, 1999) is central to the application described in this paper
because of the reasons outlined above.

Briefly, if one decomposes the convolution kernel into N basis functions
with constant shape, the problem reduces to a linear least
squares fit for the amplitudes of all kernel components. Convolutions
of the reference image with each of the N kernel components become
basis vectors and can be used to fit any other image of the same
stellar field provided it has been resampled to the coordinate grid
of the reference image. The original method assumes that there are no spatial
gradients of the PSF. Remarkably, this algorithm works best in crowded fields,
where the majority of pixels contain information about the PSF. It can even
treat some weak PSF gradients if we can afford subdividing the image into
smaller pieces.

The implementation gets complicated when the above assumption of a constant
kernel breaks down. In this case the number of coefficients needed in
order to fit spatial variation of the n-th order is $(n+1)\times(n+2)/2$
larger than in the previous problem and direct extension of the algorithm
induces unrealistic computing requirements. However, by assuming that
spatial variations of the kernel are negligible on the scale of the
kernel size, one can reduce the most time consuming calculation, that
of elements of the least squares matrix, to shuffling elements of the
corresponding (much smaller) matrix from the previous problem with constant
convolution kernel.

We found that the constant part of the kernel solution was well described
by three Gaussians with sigmas 0.7, 1.2 and 1.8 pix modified
by polynomials of orders 4, 3 and 2, respectively. We allow only the first
order spatial variation of the kernel since there are only 9 stars
that can be used in the fit. In the output we have a subtracted image with
the seeing of the currently processed frame and intensity scale corresponding
to the reference frame, plus the best fit convolution kernel. In all difference
images the residuals of constant stars were consistent with the photon noise of
a single frame. The mean of about 20 frames with the best seeing, low sky
background and resampled onto the same coordinate grid is the correct choice
of the reference image and allows approaching the photon noise limit.
We used the 18 best frames to obtain a good reference image, practically
noiseless compared to a single exposure. In difference images the galaxy
is completely removed outside the region dominated by the lens with residuals
symmetric about zero and consistent with the photon noise. We believe that
this is also the case inside this region. Fig.~1 shows central parts of the
reference image and typical test image, as well as the corresponding
difference image and the best fit PSF matching kernel. Sample difference
images from various epochs are shown in Fig.~2. Variability of all quasar
components is obvious.

The difference image contains only the variable part of the flux and can be
used for high precision relative photometry. After the galaxy is removed
remaining light can be modeled with 4 PSFs. The geometry of the lensed quasar
images is known with a very high accuracy from HST data. We adopted positions
relative to component ``A'' from Crane et al. (1991). In numerous difference
frames, ``A'' was the only quasar image which significantly changed its
brightness with respect to the reference frame. Therefore the PSF component
at the position of ``A'', corresponding to the difference flux, was practically
free of contamination by the remaining components of the blend.
This simplified calculation of its centroid in a stack of 40 such frames
and thus obtaining positions of all lensed quasar images.
The last step is a linear fit to the amplitudes of the
four lensed components in the difference image. The area dominated by variable
light and therefore useful for this purpose consists of pixels with centers
not further than 2.4 pix from any of the quasar images.
We modeled the first order spatial variation of the PSF in the reference
image using the code written for the DENIS survey (Alard 1999, in preparation).
For any other frame (and difference image) the PSF is calculated simply
by convolving the reference PSF with the best fit PSF matching kernel at the
position of the measured object.

Errors of the individual photometric points were estimated by propagating
the photon noise through the linear least squares fit and adding in quadrature
a correction for uncertain flatfield at the level of 1\% of the mean amplitude
of ``A'' component. This simple noise model gives error bars consistent with
the scatter of the individual measurements. 

\subsection{Zero point calibration}

The procedure described in the previous paragraph gives only a variable
part of the flux, for example in counts per second. Putting the light curve
on a magnitude scale requires the knowledge of the absolute amplitude
of all components and a reference star in at least one image.
In the presence of the intervening galaxy this requires a reasonably good model
of the underlying light distribution.

We used public HST images from the archive at STScI\footnotemark. The images
were taken on June 23, 1995 with the WFPC2 using F555 filter, centered on
$\lambda = 5407.0 ~ \rm \AA$, the closest available band to $V$. Proposal
ID is 5236. Four of the images had exposure times of 200 seconds and
one of them was exposed for 800 seconds.

\footnotetext{NASA/ESA Hubble Space Telescope is operated by AURA
under NASA contract NAS 5-26555}

The galaxy template was prepared by symmetrization with respect to the
brightest pixel. In the annulus contaminated by quasar images symmetric pairs
of pixels are examined and lower of the two is adopted as the best guess at
the value of both pixels with 0.6$\sigma$ correction for the bias due to
minimization. This simplified version of the symmetrization technique
used by the SDSS survey for deblending star and galaxy images
(Lupton 1999, in preparation) removes effectively quasar
components ``A'' and ``B'', however it fails for ``C'' and ``D'', due to their
very symmetric position with respect to the galaxy nucleus.
Fortunately in WFPC2 images galaxy is smooth on a much larger scale than
the area occupied by any of the quasar images. It is possible to make a local
fit to the galaxy light and subtract the faintest image before symmetrization,
which solves the problem.

The galaxy template must be rotated, resampled to the pixel grid of the OGLE
reference frame, and degraded to the seeing of each OGLE test frame before
the fit. We fitted a model consisting of 4 PSFs and the galaxy template
to every image in our data set. This is much like the conventional approach,
except that in the process we used PSF matching kernels obtained with
the image subtraction software. The scatter of this photometry was about
a factor of two larger than in the image subtraction method. Nevertheless,
the weighted mean of the difference between both light curves (in counts)
is the desired amplitude of a given quasar component in the reference image.
The statistical quality of our zero point is about 2, 3, 8 and 13 \% for
components ``A'', ``B'', ``C'' and ``D'' respectively, therefore it is
much worse than the accuracy of the difference signal with respect to the
reference frame. The data were reduced to the standard $V$ magnitudes based
on observations of standard stars from selected fields of Landolt (1992)
obtained on 11 photometric nights. We get $V=17.466\pm0.018$ and
$18.138\pm0.021$ mag for stars $\alpha$ and $\beta$ of Corrigan et al. (1991),
brighter only by about 0.038 mag compared to the photometry obtained indirectly
by these authors.

\newpage

\section{Results}

Fig.~3 shows the light curve of the QSO 2237+0305. Photometry is given in
Table~1. Machine readable data can be obtained from the OGLE web page
(see the last paragraph for the address).
The fifth order polynomial fits to light variations are also shown to guide
the eye. All components display significant variations, especially between
the two observing seasons, and even more importantly all of them varied
differently. The 18.14 mag reference star measured using exactly the same
method was constant within the errors which confirms that our photometry
is correct (Fig.~3). Component ``C'' brightened by as much as 0.7 mag and
actually almost exchanged in brightness with ``B''. The easiest explanation
to these phenomena is microlensing in the bulge of the macro lensing galaxy.
For the first time we see it happening in a sense that smooth
variation of source amplification is observed, most striking for
component ``A''. Wambsganss, Paczy\'nski and Schneider (1990) demonstrated
huge diversity of possible light variations produced by complicated network
of clustered microcaustics and showed that the sharpest features present
in the light curve are directly related to the size of the source,
the masses of the microlenses and the transverse velocity. They also
stressed that frequently sampled light curves should constrain some of
these unknowns. However in this paper we do not attempt any further
theoretical interpretation of the data.

It must be emphasized that difference photometry with respect to the
reference image is very accurate, however the overall normalization
of the light curve can be off by 15 \% for the faintest component.
On magnitude scale this affects the shape of the light curve too, since for
any test image we have:

$$ {\rm m}_V = const. - 2.5 \times
\log\left({{f_{V,\rm ref} + \Delta f_V} \over norm}\right) ~~~ \rm mag $$

In the above formula the difference flux $\Delta f_V$ is known with high
precision, however the amplitude of a given component in the reference
frame $f_{V,\rm ref}$ is known less accurately. For some applications it may
be safer to go back to the linear scale and obtain zero point free shape
of the light curve. This is accomplished for instance if we express flux
density in milli Janskys ($10^{-29}$ Wm$^{-2}$Hz$^{-1}$) :

$$ f_V = f_{V,\rm ref} + \Delta f_V =
3.67 \times 10^{(-0.4 \cdot {\rm m}_V + 6)} ~~~ \rm mJy ,$$

\noindent
as has been done in Fig.~4. 

\section{Discussion and future perspective}

We demonstrated that good quality monitoring of QSO 2237+0305 is possible with
1.3 m telescope. Despite the fact that complications characteristic of both
crowded and sparse fields are present, photometry of this system is well
handled by image subtraction method of Alard \& Lupton (1998) and Alard (1999)
supplemented with PSF fitting of the difference image. Resulting light curves
are significantly better than any other preexisting measurements, even with
much larger instruments. Probably the most important improvement is due to
much better time coverage, especially during 1997 observing season.
Beginning from 1999 we expect to obtain similar light curves during entire
period when this object is accessible from LCO. It is likely that the zero
point will be improved in the future using overlapping observations 
from instruments with better seeing and/or better galaxy template.

All data processing described in Section~3 has been integrated
to the level, at which it is possible to run reductions automatically
once a new image is collected and simply wait for the new point to be added
to a postscript plot. Photometric pipeline consists of several stand by
programs for each step of reductions, controlled by a shell script.
It is planned that this software will become a part of the OGLE photometric
pipeline at LCO. This will provide an easy way of checking the state
of Huchra's lens in real time and therefore we should be able to detect
caustic crossings early enough to issue an alert. Caustic crossing provides
in principle the way of resolving spatially the source, which in this case
would place a very tight limit on the size of the quasar
(Wambsganss, Paczy\'nski and Schneider 1990). For this measurement it is
essential to get good coverage of a High Amplification Event, with larger
instruments and better instrumental seeing if possible, therefore
the importance of early detection of HAE cannot be overestimated.
Up-to-date information on Huchra's lens is available from the OGLE web page at
http://www.astrouw.edu.pl/$\sim$ftp/ogle and its USA mirror
http://www.astro.princeton.edu/$\sim$ogle.

\acknowledgments{
We would like to thank Prof. Bohdan Paczy\'nski for
pointing to us the importance of this research, numerous useful discussions
and constant support at all stages of the OGLE project. We also thank
Robert Lupton for his insights on image processing. Comments by Dave Goldberg
and Robert Lupton helped us to improve the manuscript. This work was supported
with the Polish KBN grant 2P03D00814 to A. Udalski and
NSF grant AST--9530478 to B. Paczy\'nski.
}


\newpage

\begin{figure}[t]
\plotfiddle{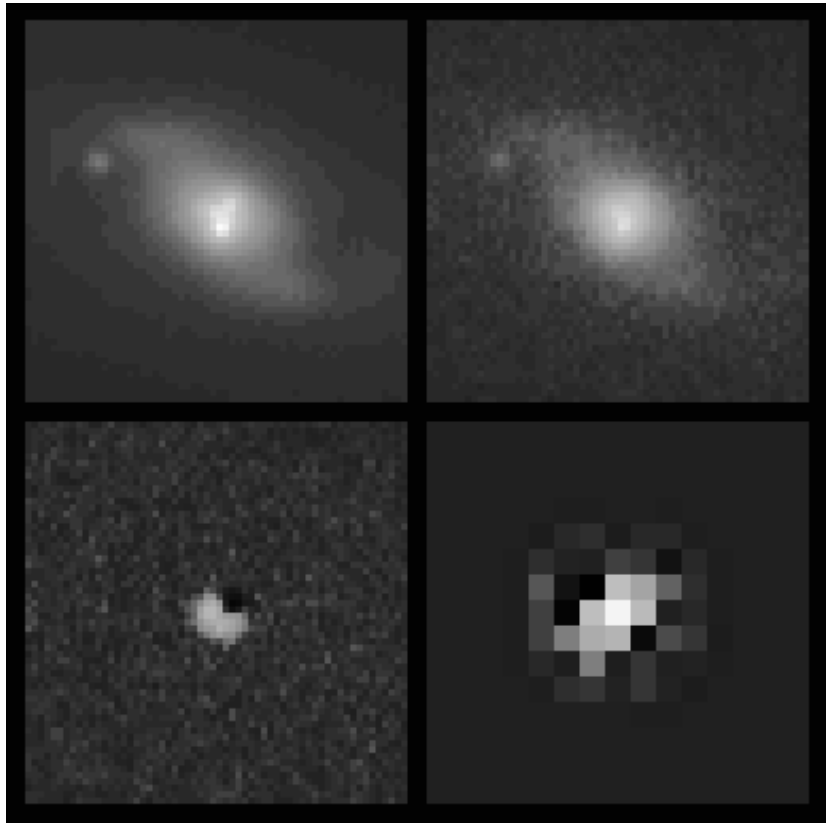}{10.5cm}{0}{80}{80}{-250}{-160}

\caption{Image subtraction as a solution to the problem of the
contamination by complicated light distribution from the lensing
galaxy. The two top panels are images of the QSO 2237+0305:
reference image and typical test image (left and right respectively).
The bottom panels show the corresponding difference image on the left
and the best fit PSF matching kernel on the right. Note that the galaxy
is completely removed by subtraction and the variable part of the flux
is well modeled by the sum of the four PSF components at known positions.
North is up and east is to the left.}

\end{figure}

\newpage

\begin{figure}[t]
\plotfiddle{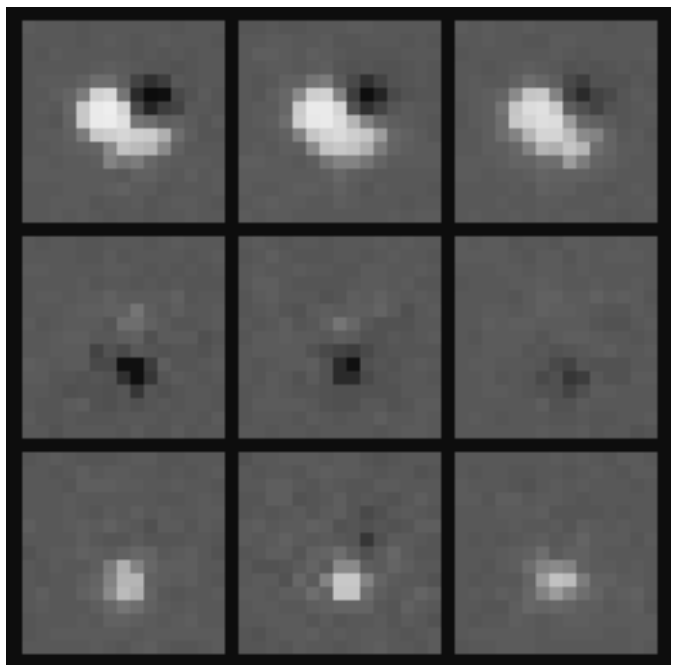}{9cm}{0}{80}{80}{-250}{-180}

\caption{Sample of nine difference images of the QSO 2237+0305 at various
epochs from the OGLE data. North is up and east is to the left.}

\end{figure}

\newpage

\begin{figure}[t]
\plotfiddle{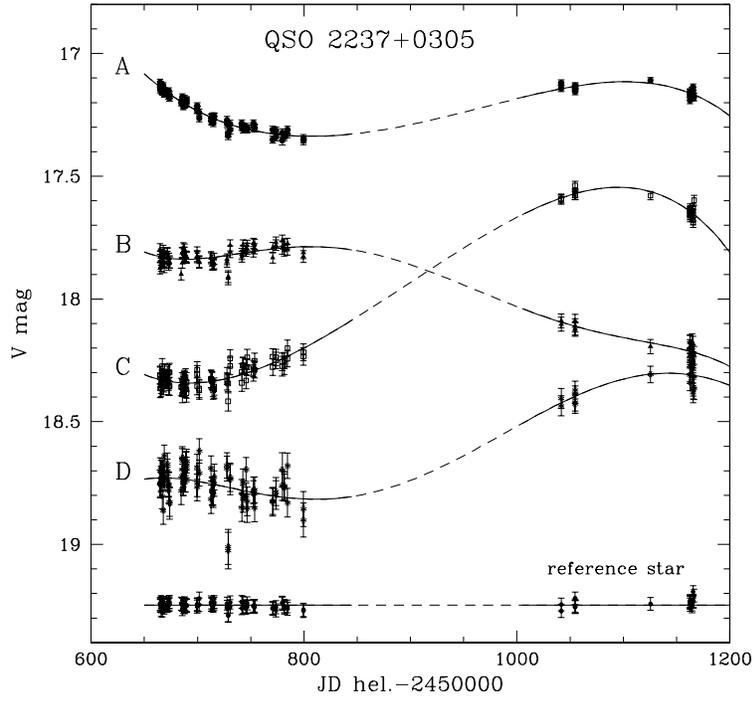}{10cm}{0}{50}{50}{-160}{-75}

\caption{Light curve of the QSO 2237+0305. Also shown is the light curve
of the 18.14 mag reference star shifted by 1.1 mag for clarity.
The polynomial fits help to assign the photometric points to each
of the components.}
\end{figure}

\begin{figure}[t]
\plotfiddle{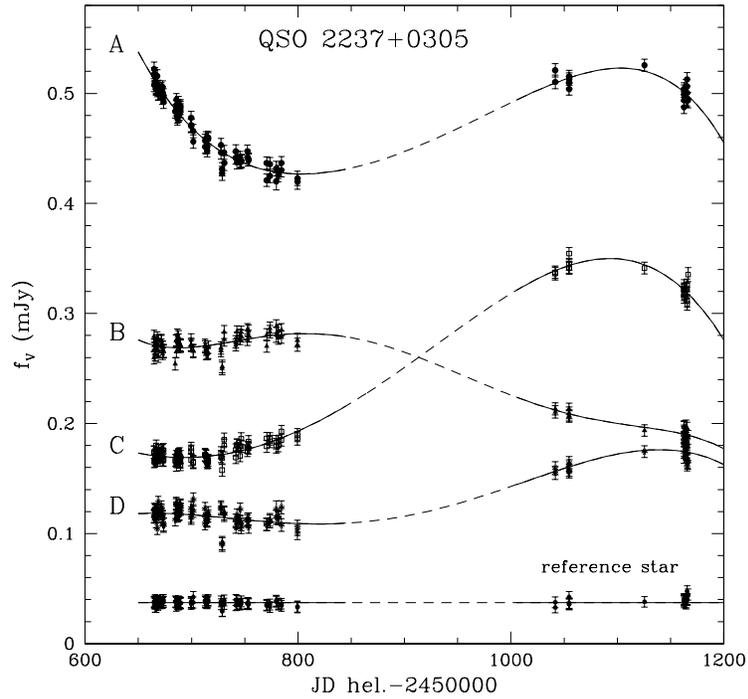}{8cm}{0}{50}{50}{-160}{-75}

\caption{Same as Fig.~3 in mJy. Please note that the amplitudes
of microlensing events in this linear plot are free of uncertainties
associated with the zero point. The flux of the reference star has been
lowered by 0.165 mJy for clarity.}

\end{figure}

\newpage

\begin{deluxetable}{lrccccccccc}
\tablecaption{Photometry of the QSO 2237+0305.}
\tablewidth{520pt}
\tablehead{
\colhead{Date} &
\colhead{[JD]\tablenotemark{\ddagger}} &
\colhead{FWHM} &
\colhead{} &
\colhead{} &
\colhead{} &
\colhead{$V$} &
\colhead{mag} &
\colhead{} &
\colhead{} &
\colhead{}
\\
\colhead{UT} &
\colhead{} &
\colhead{arc sec} &
\colhead{$A$} &
\colhead{err} &
\colhead{$B$} &
\colhead{err} &
\colhead{$C$} &
\colhead{err} &
\colhead{$D$} &
\colhead{err}
}
\startdata
97-08-04.344 &  664.849 & 1.3 & 17.127 & 0.013 & 17.829 & 0.023 & 18.310 & 0.036 & 18.754 & 0.057 \nl
97-08-04.353 &  664.858 & 1.3 & 17.148 & 0.013 & 17.797 & 0.022 & 18.313 & 0.035 & 18.752 & 0.054 \nl
97-08-04.358 &  664.863 & 1.3 & 17.117 & 0.013 & 17.874 & 0.024 & 18.364 & 0.037 & 18.701 & 0.054 \nl
97-08-05.284 &  665.789 & 1.2 & 17.143 & 0.012 & 17.865 & 0.022 & 18.339 & 0.034 & 18.737 & 0.049 \nl
97-08-05.289 &  665.794 & 1.3 & 17.131 & 0.012 & 17.825 & 0.021 & 18.351 & 0.034 & 18.705 & 0.048 \nl
97-08-05.294 &  665.799 & 1.2 & 17.127 & 0.012 & 17.840 & 0.022 & 18.363 & 0.035 & 18.731 & 0.050 \nl
97-08-06.302 &  666.808 & 1.2 & 17.156 & 0.012 & 17.827 & 0.020 & 18.274 & 0.031 & 18.803 & 0.051 \nl
97-08-06.307 &  666.813 & 1.3 & 17.152 & 0.012 & 17.840 & 0.022 & 18.325 & 0.034 & 18.698 & 0.049 \nl
97-08-06.312 &  666.818 & 1.2 & 17.154 & 0.012 & 17.829 & 0.021 & 18.322 & 0.033 & 18.705 & 0.049 \nl
97-08-07.314 &  667.819 & 1.2 & 17.158 & 0.012 & 17.825 & 0.021 & 18.324 & 0.033 & 18.712 & 0.049 \nl
97-08-07.318 &  667.824 & 1.3 & 17.130 & 0.012 & 17.802 & 0.021 & 18.357 & 0.034 & 18.862 & 0.056 \nl
97-08-08.293 &  668.799 & 1.3 & 17.150 & 0.011 & 17.871 & 0.021 & 18.297 & 0.031 & 18.721 & 0.047 \nl
97-08-08.299 &  668.804 & 1.3 & 17.166 & 0.012 & 17.810 & 0.020 & 18.347 & 0.033 & 18.638 & 0.043 \nl
97-08-08.304 &  668.810 & 1.3 & 17.166 & 0.011 & 17.849 & 0.021 & 18.318 & 0.032 & 18.731 & 0.047 \nl
97-08-11.279 &  671.785 & 1.2 & 17.159 & 0.012 & 17.814 & 0.020 & 18.331 & 0.033 & 18.699 & 0.046 \nl
97-08-11.284 &  671.790 & 1.3 & 17.166 & 0.012 & 17.810 & 0.020 & 18.330 & 0.033 & 18.673 & 0.045 \nl
97-08-11.290 &  671.796 & 1.3 & 17.155 & 0.012 & 17.827 & 0.021 & 18.306 & 0.033 & 18.734 & 0.049 \nl
97-08-12.288 &  672.794 & 1.2 & 17.163 & 0.012 & 17.860 & 0.022 & 18.338 & 0.034 & 18.754 & 0.050 \nl
97-08-12.294 &  672.800 & 1.3 & 17.174 & 0.012 & 17.835 & 0.021 & 18.356 & 0.034 & 18.758 & 0.050 \nl
97-08-12.299 &  672.805 & 1.3 & 17.152 & 0.012 & 17.835 & 0.021 & 18.300 & 0.033 & 18.780 & 0.052 \nl
97-08-13.199 &  673.704 & 1.3 & 17.181 & 0.014 & 17.852 & 0.023 & 18.356 & 0.037 & 18.827 & 0.060 \nl
97-08-13.203 &  673.709 & 1.4 & 17.169 & 0.014 & 17.849 & 0.025 & 18.299 & 0.037 & 18.833 & 0.065 \nl
97-08-24.210 &  684.716 & 1.3 & 17.201 & 0.013 & 17.899 & 0.024 & 18.359 & 0.036 & 18.782 & 0.056 \nl
97-08-25.211 &  685.717 & 1.2 & 17.197 & 0.012 & 17.812 & 0.020 & 18.389 & 0.034 & 18.655 & 0.045 \nl
97-08-25.216 &  685.722 & 1.2 & 17.191 & 0.012 & 17.811 & 0.021 & 18.333 & 0.033 & 18.647 & 0.045 \nl
97-08-25.220 &  685.726 & 1.2 & 17.177 & 0.012 & 17.852 & 0.022 & 18.361 & 0.034 & 18.722 & 0.049 \nl
97-08-26.251 &  686.757 & 1.2 & 17.215 & 0.013 & 17.793 & 0.021 & 18.346 & 0.035 & 18.716 & 0.051 \nl
97-08-26.256 &  686.762 & 1.3 & 17.211 & 0.013 & 17.833 & 0.022 & 18.352 & 0.035 & 18.748 & 0.052 \nl
97-08-26.261 &  686.767 & 1.3 & 17.197 & 0.014 & 17.811 & 0.022 & 18.339 & 0.036 & 18.747 & 0.055 \nl
97-08-27.244 &  687.750 & 1.4 & 17.205 & 0.014 & 17.799 & 0.023 & 18.327 & 0.036 & 18.697 & 0.054 \nl
97-08-27.248 &  687.754 & 1.3 & 17.209 & 0.013 & 17.811 & 0.022 & 18.330 & 0.035 & 18.733 & 0.052 \nl
97-08-27.253 &  687.759 & 1.2 & 17.183 & 0.013 & 17.829 & 0.022 & 18.360 & 0.036 & 18.734 & 0.053 \nl
97-08-28.202 &  688.708 & 1.3 & 17.192 & 0.013 & 17.828 & 0.022 & 18.366 & 0.035 & 18.671 & 0.048 \nl
97-08-28.207 &  688.713 & 1.2 & 17.206 & 0.013 & 17.826 & 0.021 & 18.318 & 0.034 & 18.735 & 0.050 \nl
97-08-29.191 &  689.697 & 1.3 & 17.199 & 0.014 & 17.832 & 0.023 & 18.338 & 0.038 & 18.676 & 0.053 \nl
97-08-29.196 &  689.702 & 1.3 & 17.193 & 0.013 & 17.829 & 0.022 & 18.384 & 0.037 & 18.661 & 0.049 \nl
97-08-29.202 &  689.708 & 1.2 & 17.189 & 0.013 & 17.829 & 0.022 & 18.369 & 0.036 & 18.694 & 0.050 \nl
97-09-08.160 &  699.666 & 1.3 & 17.230 & 0.014 & 17.810 & 0.023 & 18.306 & 0.035 & 18.751 & 0.055 \nl
97-09-08.165 &  699.671 & 1.3 & 17.214 & 0.013 & 17.826 & 0.022 & 18.289 & 0.034 & 18.767 & 0.055 \nl
97-09-08.170 &  699.676 & 1.4 & 17.229 & 0.014 & 17.830 & 0.022 & 18.355 & 0.036 & 18.684 & 0.051 \nl
97-09-10.140 &  701.646 & 1.3 & 17.241 & 0.014 & 17.851 & 0.023 & 18.372 & 0.037 & 18.709 & 0.053 \nl
97-09-10.146 &  701.652 & 1.3 & 17.264 & 0.015 & 17.848 & 0.024 & 18.365 & 0.037 & 18.619 & 0.050 \nl
97-09-21.165 &  712.671 & 1.3 & 17.275 & 0.014 & 17.829 & 0.023 & 18.327 & 0.036 & 18.816 & 0.058 \nl
97-09-21.173 &  712.678 & 1.4 & 17.262 & 0.014 & 17.834 & 0.023 & 18.325 & 0.035 & 18.690 & 0.051 \nl
97-09-22.173 &  713.679 & 1.3 & 17.257 & 0.014 & 17.843 & 0.023 & 18.335 & 0.036 & 18.781 & 0.057 \nl
97-09-22.179 &  713.685 & 1.2 & 17.276 & 0.015 & 17.858 & 0.023 & 18.333 & 0.036 & 18.785 & 0.056 \nl
97-09-23.183 &  714.689 & 1.2 & 17.276 & 0.013 & 17.859 & 0.022 & 18.373 & 0.035 & 18.798 & 0.052 \nl
97-09-23.189 &  714.695 & 1.2 & 17.284 & 0.014 & 17.861 & 0.022 & 18.363 & 0.035 & 18.788 & 0.052 \nl
97-09-24.178 &  715.684 & 1.3 & 17.255 & 0.014 & 17.855 & 0.022 & 18.357 & 0.035 & 18.752 & 0.052 \nl
97-09-24.183 &  715.689 & 1.2 & 17.258 & 0.014 & 17.836 & 0.022 & 18.368 & 0.037 & 18.725 & 0.052 \nl
97-10-06.124 &  727.629 & 1.3 & 17.271 & 0.015 & 17.848 & 0.023 & 18.314 & 0.035 & 18.678 & 0.051 \nl
97-10-06.132 &  727.637 & 1.3 & 17.288 & 0.014 & 17.838 & 0.022 & 18.336 & 0.035 & 18.691 & 0.051 \nl
97-10-07.238 &  728.744 & 1.3 & 17.335 & 0.016 & 17.916 & 0.025 & 18.344 & 0.038 & 19.025 & 0.074 \nl
97-10-07.248 &  728.753 & 1.3 & 17.325 & 0.016 & 17.909 & 0.025 & 18.417 & 0.039 & 19.010 & 0.071 \nl
97-10-09.203 &  730.708 & 1.3 & 17.288 & 0.016 & 17.782 & 0.023 & 18.273 & 0.035 & 18.727 & 0.057 \nl
97-10-09.208 &  730.713 & 1.3 & 17.311 & 0.017 & 17.808 & 0.024 & 18.242 & 0.036 & 18.736 & 0.062 \nl
97-10-20.097 &  741.601 & 1.3 & 17.285 & 0.015 & 17.817 & 0.022 & 18.297 & 0.034 & 18.792 & 0.057 \nl
97-10-20.101 &  741.605 & 1.2 & 17.300 & 0.014 & 17.828 & 0.022 & 18.342 & 0.035 & 18.850 & 0.057 \nl
97-10-21.118 &  742.622 & 1.3 & 17.310 & 0.015 & 17.781 & 0.021 & 18.272 & 0.033 & 18.745 & 0.053 \nl
97-10-21.122 &  742.626 & 1.2 & 17.298 & 0.014 & 17.803 & 0.021 & 18.270 & 0.032 & 18.814 & 0.055 \nl
97-10-24.119 &  745.623 & 1.2 & 17.299 & 0.014 & 17.805 & 0.021 & 18.276 & 0.032 & 18.695 & 0.049 \nl
97-10-24.123 &  745.627 & 1.2 & 17.310 & 0.014 & 17.792 & 0.021 & 18.332 & 0.034 & 18.804 & 0.053 \nl
97-10-25.131 &  746.635 & 1.2 & 17.308 & 0.015 & 17.796 & 0.022 & 18.237 & 0.033 & 18.825 & 0.058 \nl
97-10-25.135 &  746.639 & 1.3 & 17.306 & 0.014 & 17.782 & 0.021 & 18.279 & 0.033 & 18.852 & 0.058 \nl
97-10-31.058 &  752.562 & 1.2 & 17.285 & 0.014 & 17.773 & 0.021 & 18.283 & 0.033 & 18.780 & 0.053 \nl
97-10-31.062 &  752.566 & 1.2 & 17.298 & 0.014 & 17.808 & 0.021 & 18.302 & 0.033 & 18.793 & 0.053 \nl
97-11-01.040 &  753.544 & 1.2 & 17.303 & 0.014 & 17.777 & 0.021 & 18.252 & 0.032 & 18.777 & 0.053 \nl
97-11-01.044 &  753.548 & 1.4 & 17.304 & 0.014 & 17.796 & 0.021 & 18.289 & 0.033 & 18.852 & 0.056 \nl
97-11-18.086 &  770.588 & 1.3 & 17.311 & 0.014 & 17.791 & 0.021 & 18.278 & 0.033 & 18.823 & 0.055 \nl
97-11-18.090 &  770.592 & 1.3 & 17.351 & 0.015 & 17.832 & 0.022 & 18.235 & 0.032 & 18.828 & 0.056 \nl
97-11-21.087 &  773.589 & 1.3 & 17.314 & 0.015 & 17.781 & 0.022 & 18.251 & 0.033 & 18.785 & 0.057 \nl
97-11-21.091 &  773.593 & 1.2 & 17.341 & 0.016 & 17.769 & 0.022 & 18.238 & 0.033 & 18.798 & 0.058 \nl
97-11-27.094 &  779.595 & 1.4 & 17.324 & 0.017 & 17.763 & 0.023 & 18.256 & 0.035 & 18.760 & 0.061 \nl
97-11-27.098 &  779.599 & 1.3 & 17.353 & 0.020 & 17.791 & 0.027 & 18.274 & 0.040 & 18.696 & 0.069 \nl
97-11-29.039 &  781.540 & 1.3 & 17.332 & 0.015 & 17.777 & 0.021 & 18.251 & 0.033 & 18.769 & 0.055 \nl
97-11-29.043 &  781.544 & 1.3 & 17.339 & 0.015 & 17.803 & 0.022 & 18.274 & 0.034 & 18.761 & 0.056 \nl
97-12-02.053 &  784.554 & 1.3 & 17.328 & 0.015 & 17.774 & 0.022 & 18.200 & 0.032 & 18.680 & 0.052 \nl
97-12-02.057 &  784.558 & 1.3 & 17.311 & 0.015 & 17.799 & 0.022 & 18.244 & 0.033 & 18.830 & 0.058 \nl
97-12-17.035 &  799.534 & 1.3 & 17.354 & 0.018 & 17.811 & 0.025 & 18.234 & 0.036 & 18.856 & 0.072 \nl
97-12-17.040 &  799.539 & 1.3 & 17.346 & 0.017 & 17.827 & 0.024 & 18.218 & 0.034 & 18.901 & 0.069 \nl
98-08-16.168 & 1041.674 & 1.2 & 17.142 & 0.013 & 18.103 & 0.029 & 17.595 & 0.019 & 18.406 & 0.041 \nl
98-08-16.174 & 1041.680 & 1.3 & 17.120 & 0.013 & 18.089 & 0.028 & 17.591 & 0.018 & 18.436 & 0.041 \nl
98-08-29.224 & 1054.730 & 1.3 & 17.143 & 0.012 & 18.089 & 0.026 & 17.538 & 0.017 & 18.420 & 0.037 \nl
98-08-29.228 & 1054.734 & 1.3 & 17.136 & 0.012 & 18.125 & 0.027 & 17.578 & 0.017 & 18.429 & 0.038 \nl
98-08-29.237 & 1054.743 & 1.4 & 17.131 & 0.012 & 18.108 & 0.027 & 17.568 & 0.017 & 18.371 & 0.036 \nl
98-08-29.241 & 1054.747 & 1.3 & 17.156 & 0.012 & 18.120 & 0.027 & 17.579 & 0.017 & 18.389 & 0.036 \nl
98-11-08.069 & 1125.572 & 1.2 & 17.110 & 0.011 & 18.194 & 0.029 & 17.579 & 0.017 & 18.308 & 0.033 \nl
98-12-15.039 & 1162.539 & 1.3 & 17.192 & 0.013 & 18.177 & 0.030 & 17.645 & 0.019 & 18.206 & 0.032 \nl
98-12-15.047 & 1162.546 & 1.2 & 17.160 & 0.012 & 18.229 & 0.030 & 17.667 & 0.018 & 18.340 & 0.035 \nl
98-12-15.051 & 1162.550 & 1.2 & 17.166 & 0.012 & 18.195 & 0.029 & 17.640 & 0.018 & 18.311 & 0.034 \nl
98-12-15.057 & 1162.556 & 1.2 & 17.178 & 0.012 & 18.228 & 0.030 & 17.659 & 0.018 & 18.250 & 0.032 \nl
98-12-15.061 & 1162.560 & 1.2 & 17.155 & 0.012 & 18.175 & 0.028 & 17.630 & 0.018 & 18.272 & 0.032 \nl
98-12-17.037 & 1164.536 & 1.3 & 17.150 & 0.013 & 18.173 & 0.031 & 17.643 & 0.019 & 18.284 & 0.037 \nl
98-12-17.041 & 1164.540 & 1.3 & 17.178 & 0.014 & 18.221 & 0.033 & 17.665 & 0.020 & 18.304 & 0.039 \nl
98-12-18.029 & 1165.528 & 1.3 & 17.137 & 0.013 & 18.267 & 0.034 & 17.679 & 0.020 & 18.357 & 0.039 \nl
98-12-18.033 & 1165.532 & 1.3 & 17.176 & 0.013 & 18.249 & 0.032 & 17.641 & 0.019 & 18.277 & 0.035 \nl
98-12-18.043 & 1165.542 & 1.4 & 17.163 & 0.013 & 18.183 & 0.031 & 17.688 & 0.020 & 18.382 & 0.040 \nl
98-12-18.047 & 1165.546 & 1.4 & 17.151 & 0.014 & 18.257 & 0.034 & 17.630 & 0.020 & 18.367 & 0.041 \nl
98-12-19.033 & 1166.532 & 1.3 & 17.178 & 0.015 & 18.229 & 0.036 & 17.598 & 0.021 & 18.364 & 0.046 \nl

\enddata
\tablenotetext{\ddagger}{Heliocentric $\rm JD - 2450000$}
\end{deluxetable}


\begin{references}

\reference{} Alard, C., \& Lupton, R. H., 1998, ApJ, 503, 325

\reference{} Alard, C., 1999, A\&A, submitted (= astro-ph/9903111)

\reference{} Corrigan, R. T., et al. 1991, AJ, 102, 34

\reference{} Crane, P., et al. ApJ, 369, L59

\reference{} Huchra, J., Gorenstein, M., Kent, S., Shapiro, I., Smith, G.,
Horine, E., \& Perley., R., 1985, AJ 90, 691

\reference{} Irwin, M. J., Webster, R. L., Hewett, P. C., Corrigan, R. T.,
\& Jedrzejewski, R. I., 1989, AJ, 98, 1989

\reference{} Landolt, A. U., 1992, AJ, 104, 372

\reference{} \char31 stensen, R., et al. 1996, A\&A, 309, 59

\reference{} Paczy\'nski, B., 1996, ARAA, 34, 419

\reference{} Schneider, D. P., Turner, E. L., Gunn, J. E., Hewitt, J. N.,
Schmidt, M., \& Lawrence, C. R., 1988, AJ, 95, 1619

\reference{} Udalski, A., Kubiak, M., \& Szyma\'nski, M., 1997, Acta Astron.,
47, 319

\reference{} Wambsganss, J., Paczy\'nski, B., \& Schneider, P.,
1990, ApJ, 358, L33

\reference {} Webster, R. L., Ferguson, A. M. N., Corrigan, R. T.,
\& Irwin, M. J., 1991, AJ, 102, 1939

\end{references}
\end{document}